\documentclass[10pt,preprint]{emulateapj}

\shorttitle{BIPOLAR OUTFLOW FROM THE PROTOSTELLAR CORE W51e2-E}
\shortauthors{SHI, ZHAO \& HAN }

\begin{document}

\title{A bipolar outflow from the massive protostellar core W51e2-E}
\author{Hui Shi\altaffilmark{1},
        Jun-Hui Zhao\altaffilmark{2},
        J.~L. Han\altaffilmark{1}}
\begin{abstract}

We present high resolution images of the bipolar outflow from W51e2,
which are produced from the Submillimeter Array archival data observed
for CO(3-2) and HCN(4-3) lines with angular resolutions of
$0.''8\times0.''6$ and $0.''3\times0.''2$, respectively. The images
show that the powerful outflow originates from the protostellar core
W51e2-E rather than from the ultracompact H$_{\rm II}$ region
W51e2-W. The kinematic timescale of the outflow from W51e2-E is about
1000\,yr, younger than the age ($\sim$5000\,yr) of the ultracompact
H$_{\rm II}$ region W51e2-W.  A large mass-loss rate of
$\sim1\times10^{-3}$M$_{\sun}$~yr$^{-1}$ and a high mechanical power
of 120\,$L_{\sun}$ are inferred, suggesting that an O star or a
cluster of B stars are forming in W51e2-E. The observed outflow
activity along with the inferred large accretion rate indicates that
at present W51e2-E is in a rapid phase of star formation.
\end{abstract}

\altaffiltext{1}{National Astronomical Observatories, Chinese Academy
  of Sciences, 20A DaTun Road, Beijing 100012, China;
  shihui @ nao.cas.cn, hil @ nao.cas.cn}
  \altaffiltext{2}{Harvard-Smithsonian Center for Astrophysics, 60
  Garden Street, Cambridge, MA 02138, USA; jzhao @ cfa.harvard.edu}

\section{INTRODUCTION}

Molecular outflows from protostellar cores provide a critical means to
transport angular momentum from the accretion disk into the
surrounding environment. The molecular outflows and infalls vigorously
affect the turbulence and dissipation of molecular gas in molecular
cores, playing an important role in the formation and evolution of
massive stars.  Observations of outflows can reveal the history of
mass-loss processes in a protostellar system. In particular, high
angular resolution observations of outflows can precisely determine
the origin of an outflow and separate the outflow from the infall in a
protostellar core. Nearly half of the observed molecular outflows are
driven by massive protostars ($L_{\rm bol}>10^3L_{\sun}$) and have a
typical dynamical age of $\sim10^4$\,yr and a mass outflow rate of
$\sim10^{-3}$\,M$_{\sun}$~yr$^{-1}$ \citep[e.g.][]{Chu02}.

W51e2, located at a distance of 5.1\,kpc \citep{XRM09}, is a
prototypical core for massive star formation. The molecular core is
associated with an ultracompact H$_{\rm II}$ region
\citep[e.g.][]{Sco78,GJW93} and a possible inflow \citep[e.g.][]{HY96,
  ZHO98, SZH04, SZH10}. With high-resolution observations using the
Submillimeter Array (SMA)\footnote{
The Submillimeter Array is a joint project between the Smithsonian
Astrophysical Observatory and the Academia Sinica Institute of
Astronomy and Astrophysics and is funded by the Smithsonian
Institution and the Academia Sinica.  }
at the wavelengths of 0.85 and 1.3\,mm, W51e2 was resolved into four
sub-cores \citep{SZH10} including the ultracompact H$_{\rm II}$ region
(W51e2-W) and a massive protostellar core (W51e2-E). From the analysis
of the HCN(4-3) absorption line, \citet{SZH10} showed that the bright
dust core, W51e2-E, $0.''9$ east of W51e2-W, dominates the mass
accretion in the region. A bipolar outflow has been detected in the
W51e2 region \citep{KK08} based on the SMA observations of CO(2-1)
line with an angular resolution of $\sim 1\arcsec$. However, the
resolution of the CO(2-1) observation was not adequate to determine
the origin of the molecular outflow.

High angular resolution observations are imperative to identify the
origin of the molecular outflow. In this Letter, we present
high-resolution images of the bipolar outflow using the SMA data of
the HCN(4-3) and CO(3-2) lines observed at 0.85 and 0.87\,mm with
angular resolutions of $0.''3\times0.''2$ and $0.''8\times0.''6$,
respectively. We determine and discuss the physical properties of
W51e2-E and W51e2-W cores, and assess the roles of the outflow in the
process of massive star formation in W51e2.

\begin{figure*}[t]
\centering
  \includegraphics[width=83mm,angle=-90]{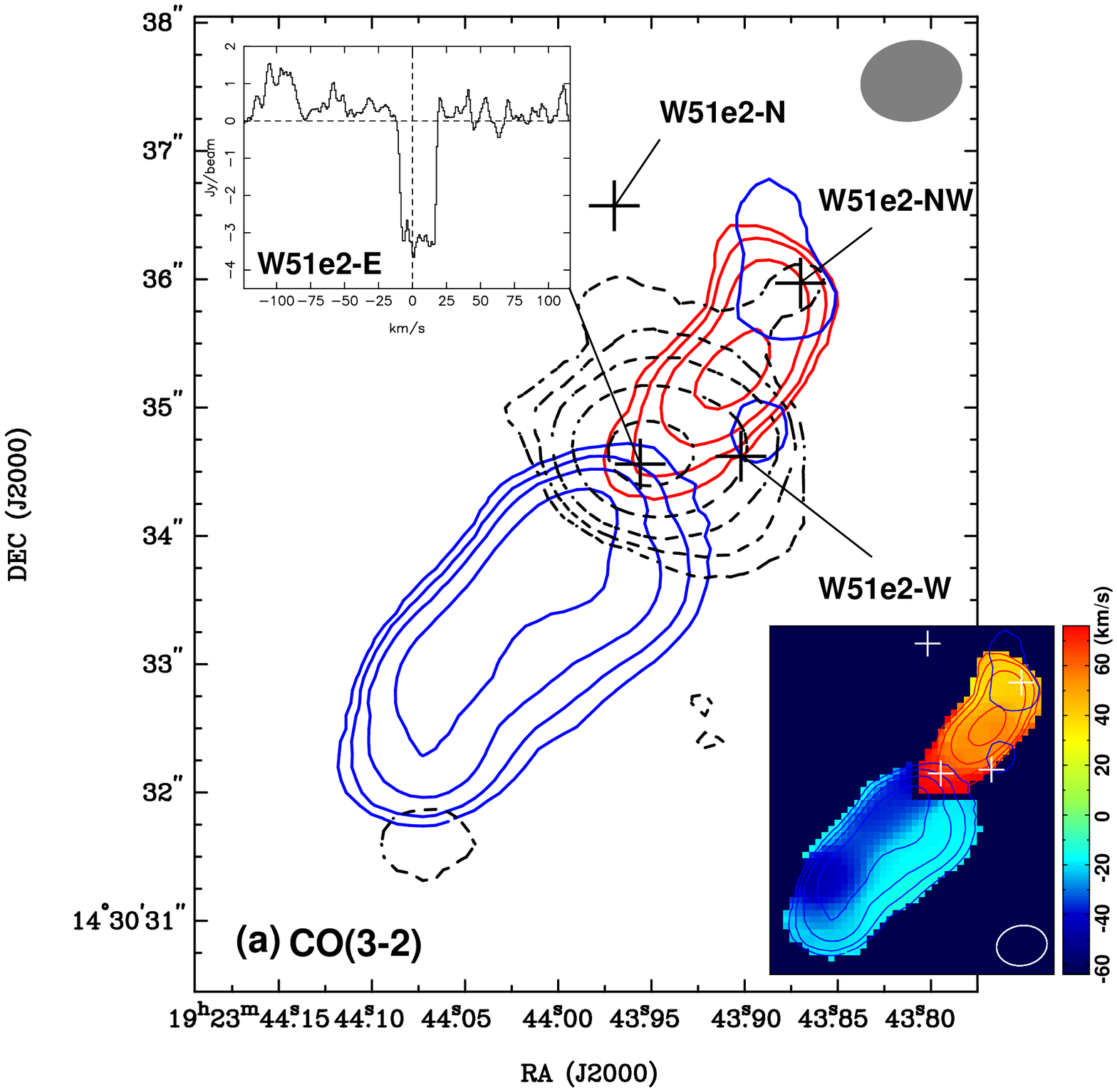}
  \includegraphics[width=83mm,angle=-90]{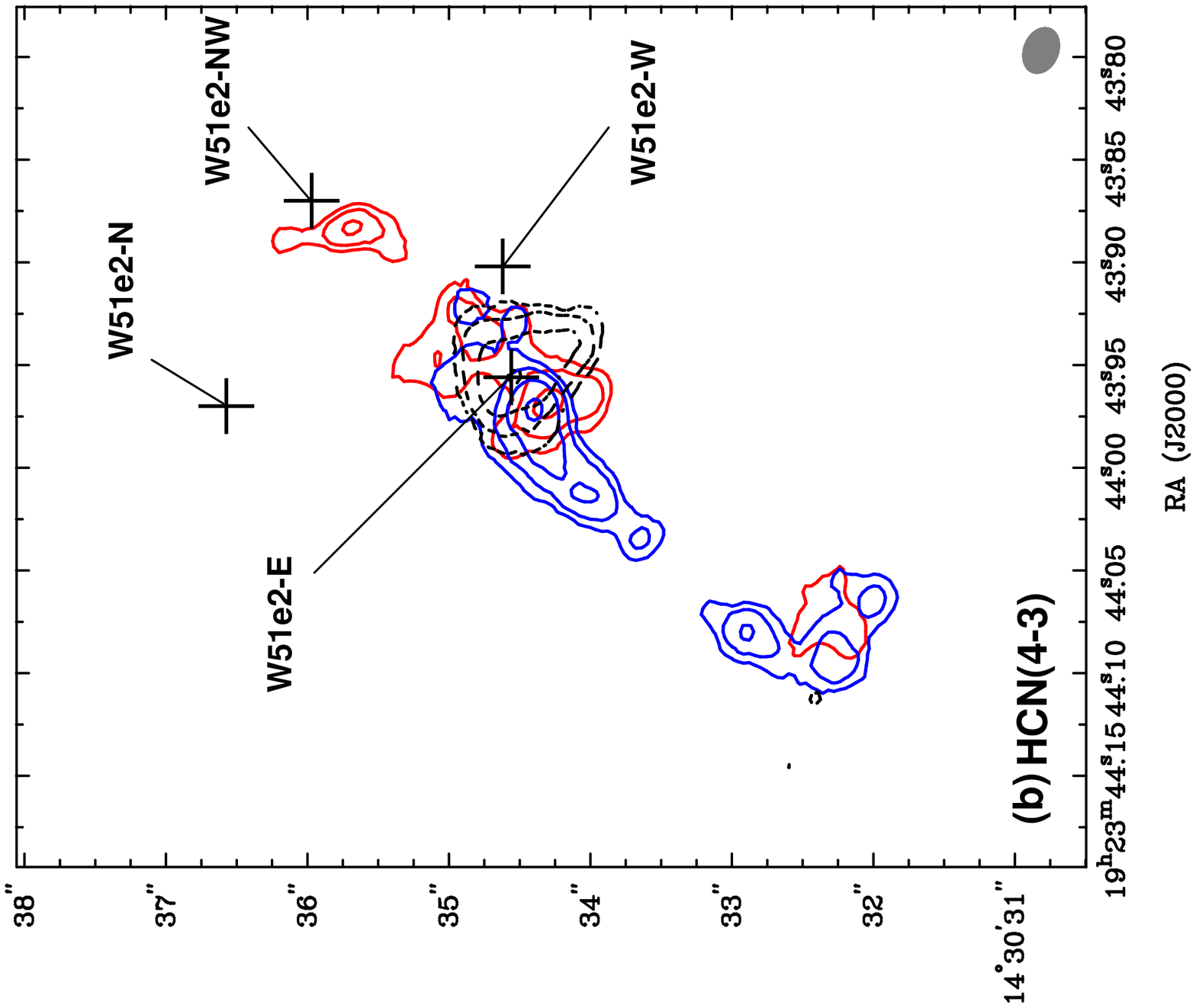}
\caption{
(a) SMA image of W51e2 for CO(3-2) line emission at 0.87\,mm,
  integrated from the velocity ranges of $-124$ to $-12$~km~s$^{-1}$
  (blue contours) and +10 to +116~km~s$^{-1}$ (red contours).  The
  integrated absorption (dashed contours) corresponds to the velocity
  range $-11$ to +9~km~s$^{-1}$. The contours are $\pm5\sigma \times
  2^n$ Jy~beam$^{-1}$~km~s$^{-1}$ ($n=0$, 1, 2, 3, ...), where
  $\sigma=$1.2, 1.3, and 0.63~Jy~beam$^{-1}$~km~s$^{-1}$ for the
  blue-, red-shifted emission, and absorption, respectively. The FWHM
  beam is $0.''79\times0.''62$ (P.A.=$-80\arcdeg$) shown at top right.
  The top left inset is the CO(3-2) absorption spectrum toward
  W51e2-E.  The bottom right inset is the intensity-weighted velocity
  image derived from the CO(3-2) line data (the color wedge
  representing the velocity range $-64$ to +76\,km~s$^{-1}$ relative to
  V$_{\rm LSR}$=53.9 km~s$^{-1}$), excluding the absorption and
  insignificant emission ($<3\sigma$).  Crosses mark the positions of
  the sub-cores in W51e2 \citep{SZH10}.
(b) SMA image of W51e2 for HCN(4-3) line at 0.85\,mm, integrated from
  the same velocity range as CO(3-2) in (a).  The contours are
  $\pm5\sigma \times 2^n$ Jy~beam$^{-1}$~km~s$^{-1}$ ($n=0$, 1, 2, 3,
  ...), where $\sigma=$0.71, 0.67, and 0.31 Jy~beam$^{-1}$~km~s$^{-1}$
  for the blue-, red-shifted emission, and absorption,
  respectively. The FWHM beam is $0.''33\times0.''24$
  (P.A.=$68\arcdeg$) shown at bottom right.
}
\label{fig:COline}
\end{figure*}

\section{Data reduction}
The interferometer data for the HCN(4-3) and CO(3-2) lines at $\nu_0=$
354.505 and 345.796~GHz were acquired from the SMA archive, observed
on 2007 June 18 and 2008 July 13 with the ``very extended'' and
``extended'' array configurations, respectively.  The reduction for
the HCN(4-3) line data was made in Miriad \citep{STW95} following the
reduction instructions for SMA data\footnote{\it
  http://www.cfa.harvard.edu/sma/miriad}, which has been discussed in
\citet{SZH10}.  We constructed image cubes for the HCN(4-3) line with
a channel width of 1 km s$^{-1}$ and an FWHM beam of
$0.''33\times0.''24$ (P.A.=68$\arcdeg$).  The typical rms noise in a
channel image is 0.07 Jy~beam$^{-1}$.

The CO(3-2) line is included in the SMA observations at 0.87\,mm for
the polarization study of W51e2 that was discussed by
\citet{THK09}. The cross-polarization data were flagged before
processing the data in Miriad. The data scans on 3C454.3 were used for
bandpass calibration. QSO J$1751+096$ was used for calibration of the
complex gains, and Uranus was used to determine the flux-density
scale. The continuum level was subtracted from the line channels using
UVLIN. The continuum-free line data were used for imaging with the
natural weighting and a channel width of 1~km~s$^{-1}$. A typical rms
noise of 0.14 Jy~beam$^{-1}$ and an FWHM beam of $0.''79\times0.''62$
(P.A.=$-80\arcdeg$) were achieved for a channel image.

The shortest UV lengths in the visibility data are 30 k$\lambda$ and
54 k$\lambda$, corresponding to angular sizes of about 7$\arcsec$ and
4$\arcsec$ for the CO and HCN structure, respectively. The compact
structure ($<$3\arcsec) of the outflow in both CO and HCN appears to
be adequately sampled in the SMA observations. However, the total line
flux density of the outflow may be underestimated due to missing short
spacing. From a visibility model for the outflow source with a $uv$
coverage identical to the SMA data, we assess that $\leq$2\% and
$\leq$8\% in the CO and HCN line flux densities may be underestimated,
respectively.

\section{Results}
\subsection{Morphologies of the bipolar outflow}
Figure \ref{fig:COline}(a) shows the CO(3-2) image of W51e2 for the
integrated blue-shifted ($-$124 to $-$12~km~s$^{-1}$) and red-shifted
(+10 to +116~km~s$^{-1}$) emission and absorption ($-$11 to
+9~km~s$^{-1}$) with respect to the systemic velocity of $V_{\rm
  LSR}=$ 53.9\,km~s$^{-1}$ \citep{SZH10}. The outflow velocities
discussed throughout the Letter are with respect to this systemic
velocity. The low velocity limits were determined by the velocities at
the boundary of the absorption region. The high velocity limits
correspond to the velocities at which the high-velocity wing profiles
drop to the 1$\sigma$ level.  The ranges of velocities for the outflow
wings are also valid for the CO(2-1) and HCN(4-3) line profiles. The
CO(3-2) line profiles appear to be contaminated by three narrow
($\Delta V_{\rm FWHM}\approx6$ km s$^{-1}$) anonymous lines at the
velocities of --104, --89 and +70 km s$^{-1}$. Prior to integrating
the CO(3-2) line from the outflow, the line emission in the velocity
ranges from --117 to --79 and from +62 to +76 km s$^{-1}$ is blanked.
Thus, about 20\% and 10\% of the line fluxes are missing in the blue-
and red-shifted outflow lobes shown in Figure \ref{fig:COline}(a),
respectively.

The CO(3-2) image shows a stronger blue-shifted lobe in the southeast
with a position angle of $140\degr\pm6\degr$ (from north to east) and
a weaker red-shifted lobe in northwest ($-40\degr\pm11\degr$), which
are in agreement with the outflow structure revealed in the CO(2-1)
image \citep{KK08}. With Gaussian fitting, we determined the intrinsic
sizes of $2.''3\times0.''6$ and $1.''1\times0.''6$ for the blue- and
red-shifted lobes, respectively. Our high resolution CO(3-2) image
clearly shows that the molecular bipolar outflow originates from
W51e2-E (the protostellar core) rather than W51e2-W (the ultracompact
H$_{\rm II}$ region).  The bright emission ridge along the northeast
edge of the blue-shifted lobe corresponds to a high-velocity component
of the outflow, or an outflow jet (see the velocity distribution image
in the lower right inset of Figure \ref{fig:COline}(a)). The weak,
relatively extended emission along the southwest edge of the
blue-shifted lobe may indicate a backflow of the molecular gas from
the interaction between the outflows and the medium surrounding the
W51e2 core. In addition, a velocity gradient is present along the
major axis of the red-shifted lobe, with higher velocities near the
core, indicating that the gas in the red-shifted outflow is
decelerated. By averaging the intensity-weighted velocities along the
bright emission ridges in both the blue- and red-shifted lobes, we
infer the mean radial velocities of $\bar{V}_{\rm blue}=-34\pm7$ and
$\bar{V}_{\rm red}=+52\pm2$ km s$^{-1}$ for the outflow lobes.  The
CO(3-2) spectrum toward W51e2-E also shows a large fraction of
absorption to be red-shifted relative to the systemic velocity (top
left inset of Figure \ref{fig:COline}(a)), suggesting the presence of
infalling gas, which agrees with the results from the observations of
the CO(2-1) and HCN(4-3) lines \citep{SZH10}.

\begin{table*}[t]
\small
\centering
\tablewidth{0pt}
\setlength{\tabcolsep}{7.5mm}
%
\caption{Outflow Parameters from CO(3-2) and CO(2-1) Lines.}
\label{tab:CO}
\begin{tabular}{lcccc}
\hline
\hline
Quantities & \multicolumn{2}{c}{CO(3-2)}&\multicolumn{2}{c}{CO(2-1)}  \\
         &  Blue&  Red &  Blue& Red \\
\hline
$V$ range (km~s$^{-1}$)&  --124 to --12& +10 to +116& --124 to --12& +10 to +116 \\  
$\bar{V}$  (km~s$^{-1}$)&    $-34\pm7$&     $+52\pm2$& ... & ... \\
$\Delta V$ (km~s$^{-1}$)$^{\rm a}$&  19$\pm$5 & 33$\pm$8 & 14$\pm$2    & 19$\pm$4     \\
Outflow intrinsic size $\theta_{\rm Maj}$ (\arcsec)& $2.3\pm0.3$ & $1.1\pm0.4$& ... & ... \\
Outflow intrinsic size $\theta_{\rm Min}$ (\arcsec)& $0.6\pm0.5$ & $0.6\pm0.6$& ... & ... \\
Outflow P.A. (\arcdeg)          & $140\pm6$   & $-40\pm11$ & ... & ... \\
$\int \Delta I(v){\rm d}v$ (Jy~beam$^{-1}$~km~s$^{-1}$)$^{\rm a}$ &
 $120_{-18}^{+30}$&$80_{-12}^{+15}$&$59\pm6$&$25\pm3$\\
$\Delta I(\nu_{\rm p})$  (Jy~beam$^{-1}$)$^{\rm a}$ & 6.5$\pm$1.0  &2.4$\pm$0.4 & 4.3$\pm$0.4     & 1.3$\pm$0.2 \\
$\Delta T_{\rm L}(\nu_{\rm p})$ (K)$^{\rm a, b}$     & 112$\pm$17    &41$\pm$7 &$>100$  & $>30$ \\
$f_{\rm B}^{\rm b}$              & 0.6          & 0.6        & $\le1$            &  $\le1$       \\
$T_{\rm ex}$ (K)         & 120  & 65    & 120   & 65  \\
$\tau_{\rm L}(\nu_{\rm p})$           & 3.9    & 1.3   & 1.9    &0.7   \\
$N_{\rm H_2}$ ($10^{22}$cm$^{-2}$)$^{\rm c}$   &   6.4  &  1.3   &  ...&  ... \\
$M_{\rm out}$ (M$_{\sun}$)&  1.3  & 0.1  &  ...&  ... \\
$t_{\rm out}$ (yr)$^{\rm d}$& 1600/tan($i$) & 500/tan($i$) & ...& ...\\
\hline
%
\multicolumn{5}{l}{\bf Notes:} \\
\multicolumn{5}{l}{$^{\rm a}$The quantities at the
maximum positions of the CO(3-2) flux. $\Delta V = \int \Delta I(v) {\rm d}v / \Delta I(\nu_p)$. } \\
\multicolumn{5}{l}{$^{\rm b}$The beam filling factor ($f_{\rm B}$) 
is the ratio of the minor-axis size to the beam size projected on the minor axis and  } \\
\multicolumn{5}{l}{$\Delta T_{\rm L}(\nu_{\rm p}) = \frac{c^2}{2k}\nu_{\rm p}^{-2} 
\Delta I(\nu_{\rm p})f_{\rm B}^{-1}$. } \\
\multicolumn{5}{l}{$^{\rm c}$${N_{\rm H_2} = 8\times10^{17}T_{\rm ex}{\rm
exp}\left(\frac{16.6}{T_{\rm ex}}\right)\left[{1-{\rm exp}(\frac{-16.6}{T_{\rm ex}})}\right]^{-1}
\int \tau_L {\rm d}v}$ $\approx 5\times10^{16} {\rm cm^{-2}} T_{\rm ex}^2 
{\rm exp}\left(\frac{16.6}{T_{\rm ex}}\right)
\tau_L(\nu_{\rm p})\Delta V$ for CO(3-2)}\\ 
\multicolumn{5}{l}{ assuming an abundance of ${\left[N_{\rm CO}\right]/\left[N_{\rm H_2}\right]=10^{-4}}$.}\\
\multicolumn{5}{l}{$^{\rm d}$$i$ is the angle between the major axis 
of outflow and line of sight.}\\  
\end{tabular}
\end{table*}

The HCN(4-3) image in Figure \ref{fig:COline}(b) shows the
high-density gas components in the bipolar outflow. At the resolution
of $0.''3\times0.''2$, the SMA image unambiguously verifies that
W51e2-E is the driving source for the bipolar outflow in the
region. We also noticed that the HCN emission traces the northeast
edge of the blue-shifted lobe where the CO(3-2) emission shows a sharp
gradient both in intensity and velocity, suggesting that the line
emission of the molecular HCN is likely enhanced near the shocked or
compressed regions of an outflow \citep{BP97}. In addition, at the
southeastern tip of the blue-shifted outflow lobe, an ``L'' shape
structure is detected in the HCN(4-3) emission with a double spectral
feature at $\rm V \approx -14$ and +16 km s$^{-1}$, each of the
spectral profiles having a line width of $\Delta V_{\rm FWHM}\sim 15$
km~s$^{-1}$. The HCN line emission suggests that the gas in the region
is swept up by the suspected bow shock at the tip of the outflow.

The polarization emission detected in the W51e2 region shows a
``hourglass'' shape in magnetic fields, which is probably related to
the gas infall and disk rotation \citep{THK09}. Our images for the
bipolar outflow in both the CO(3-2) and HCN(4-3) lines (Figures
\ref{fig:COline}(a) and (b)) show that, located near the ``hourglass''
center, W51e2-E appears to be the energy source responsible for both
the magnetic fields and the bipolar outflow.  The ``hourglass''
structure appears to slightly extend along the major axis of the
outflow, suggesting that the magnetic fields might also be tangled
with the outflow motions.

\subsection{Physical properties}

The bipolar outflows from W51e2-E have been detected in both CO(3-2)
and CO(2-1) with the SMA. In order to assess the physical conditions
of the outflow gas, we re-processed the CO(2-1) data and convolved the
CO(3-2) image to the same beam size as that of the CO(2-1) image
($1.''4\times0.''7$). For both the CO(3-2) and CO(2-1) lines, the line
peak ($\nu=\nu_{\rm p}$) brightness temperatures $\Delta T_L(\nu_{\rm
  p})$ of the blue- and red-shifted line profiles can be determined
from the maximum positions of the CO(3-2) line fluxes in Figure
\ref{fig:COline}. Then, the optical depth at the line peak can be
estimated if the gas is in local thermal equilibrium (LTE)
\citep{RW04} and the cosmic microwave background (CMB) radiation is
negligible,
   \begin{eqnarray}
      \tau_L(\nu) =  
      -{\rm ln}\left[1-\frac{\Delta T_L(\nu)}{T_0(\nu)}
        \left({\rm e}^{\frac{T_0(\nu)}{T_{\rm ex}}}
        -1\right)\right],
   \label{eq:tau_nu}
   \end{eqnarray}
where $T_{\rm ex}$ is the excitation temperature of the transition and
$T_0(\nu)=h\nu/k$.  Thus, each of $\tau(\nu)_{(3-2)}$ and
$\tau(\nu)_{(2-1)}$ can be determined independently from Equation
(\ref{eq:tau_nu}) at a given $T_{\rm ex}$. On the other hand, under
the assumption of LTE, the optical-depth ratio between CO(3-2) and
CO(2-1) can be independently expressed as
   \begin{eqnarray}
      \frac{\tau_L(\nu)_{(3-2)}}{\tau_L(\nu)_{(2-1)}}
           &=& \frac{(S\mu^2)_{(3-2)}}{(S\mu^2)_{(2-1)}}
               \frac{{\rm e}^\frac{T_0(3-2)}{T_{\rm ex}}-1}
                    {{\rm e}^\frac{T_0(2-1)}{T_{\rm ex}}-1}
               {\rm e}^{-\frac{T_0(3-2)}{T_{\rm ex}}},
   \label{eq:tau_ratio}
   \end{eqnarray}
where $S$ and $\mu$ are the line strength and the permanent electric
dipole moment, respectively. For the two given transitions, such as
CO(3-2) and CO(2-1), the ratio of the optical depth ($\displaystyle
\tau_{L(3-2)}/\tau_{L(2-1)}$) depends only on the excitation
temperature $T_{\rm ex}$ (Equation (\ref{eq:tau_ratio})).  By giving
an initial value of $T_{\rm ex}$, the three unknown quantities
($\tau_{L(3-2)}$, $\tau_{L(2-1)}$, and $T_{\rm ex}$) can be solved
with a few iterative steps from Equations (\ref{eq:tau_nu}) and
(\ref{eq:tau_ratio}), constrained by the measured quantities $\Delta
T_{L(3-2)}$ and $\Delta T_{L(2-1)}$ for each of the blue- and
red-shifted lobes.

Table \ref{tab:CO} summarizes the results of the physical parameters
determined for the molecular outflow. For the quantities directly
determined from the CO images, the estimated 1$\sigma$ uncertainties
are given.  The excitation temperatures of 120 and 65 K are derived
for the blue- and red-shifted CO outflow lobes, respectively. The
optical depths of the blue-shifted CO lines are $\tau_{L(3-2)}=$3.9
and $\tau_{L(2-1)}=$ 1.9, and the red-shifted lobe are
$\tau_{L(3-2)}=$1.3 and $\tau_{L(2-1)}=$ 0.7. Uncertainties of
$\sim$30\% in the derived values for $T_{\rm ex}$ and $\tau_{L}$ are
mainly due to the uncertainties in the determination of the line flux
densities.  The H$_2$ column density (6.4$\times10^{22}$ cm$^{-2}$) in
the blue-shifted lobe appears to be a few times greater than the value
derived from the red-shifted lobe ($1.3\times10^{22}$ cm$^{-2}$).
Considering the intrinsic size of the outflow, we infer that the total
outflow mass is about 1.4 M$_{\sun}$.  From the intrinsic sizes
($\theta_{\rm Maj}$) and the intensity-weighted mean velocity ($\bar
V$) of the outflow components, the kinematic timescale ($t_{\rm out}$)
is found to be $\sim$1000 yr if the outflow inclination angle is taken
as $i\sim45$\arcdeg.  Therefore, the mass-loss rate and momentum rate
are 1$\times10^{-3}$ M$_{\sun}$ yr$^{-1}$ and 0.04 M$_{\sun}$ km
s$^{-1}$ yr$^{-1}$ from W51e2-E, respectively. The mass-loss rate has
the same order of magnitude as the value of the accretion rate
\citep[$> 1\times 10^{-3}$ M$_{\sun}$~yr$^{-1}$;][]{SZH10} derived
from the HCN(4-3) absorption line toward W51e2-E. The momentum rate
corresponds to a mechanical power of 120 $L_{\sun}$, which is at least
an order of magnitude greater than that of an early type B star
\citep[e.g.][]{Chu99,ASG+07}, suggesting that the protostellar core
W51e2-E is forming an O type star or a cluster of B type stars.

\section{Discussions}

\begin{figure}[htp!!!]
\centering
\includegraphics[width=90mm, angle=-90]{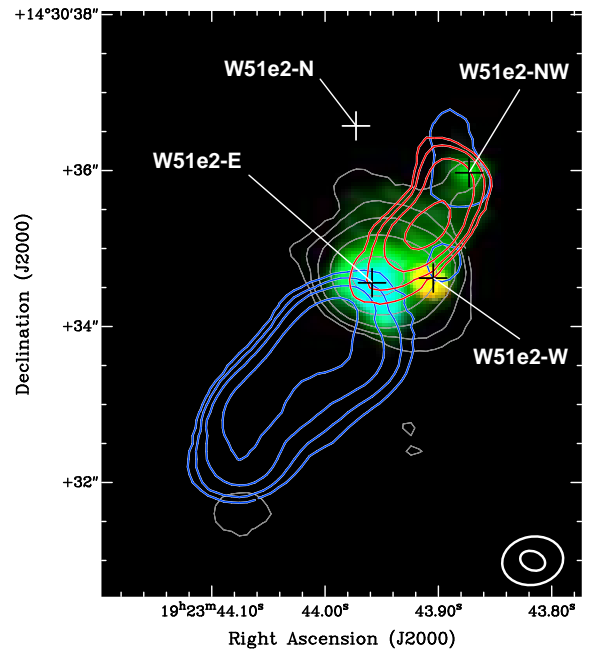}
\caption{
RGB composition image of H26$\alpha$ emission (red), 0.85 mm continuum
emission (green), and HCN(4-3) absorption (blue) observed with the SMA
at a resolution of $0.''33\times0.''24$
\citep[P.A.=68$\arcdeg$;][]{SZH10} overlaid with the contour map of
CO(3-2) outflow (red and blue) and absorption (grey) (Figure
\ref{fig:COline}a). Crosses mark the positions of the sub-cores.
}
\label{fig:RGB}
\end{figure}

The current star formation activities in the W51e2 complex are shown
by the powerful bipolar outflow originating from W51e2-E, in contrast
to the ultracompact H$_{\rm II}$ region W51e2-W detected at 6~cm
\citep{Sco78}.  The central region of W51e2-W ($<0.06\arcsec$),
characterized by a hypercompact H$_{\rm II}$ component (see the
H26$\alpha$ line emission region in Figure \ref{fig:RGB}) with
$T_e=12,000$ K, is surrounded by an expanding ionized component with a
mean temperature of 4900\,K \citep{SZH10}. The mean density of
$3.0\times10^5$\,cm$^{-3}$ and a Lyman photon rate of
$3.0\times10^{48}$\,s$^{-1}$ were also derived. Using a classic model
for the expansion of an H$_{\rm II}$ region \citep{GL99}, we estimated
the dynamic age of the H$_{\rm II}$ region as the time of the sound
wave traveling from the initial ionization front at a radius close to
the Str$\ddot{\rm o}$mgren radius \citep[$r_s$;][]{Str39} to the
current ionization front at a radius of $r_i$:
   \begin{eqnarray}
      t  =  \frac{4}{7} \left[\left(\frac{r_i}{r_s}\right)^{7/4}-1\right]
            \frac{r_s}{C_{\rm II}}.
   \label{eq:HIIage}
   \end{eqnarray}
Here $r_i$ can be determined from radio continuum observations of the
H$_{\rm II}$ region, we adopted the size of $< 2.5\arcsec$ which
corresponds to $r_i< 0.03$\,pc from the result of 3.6\,cm continuum
observation by \citet{Meh94}. We derived the Str$\ddot{\rm o}$mgren
radius of 0.008\,pc using the mean physical parameters inferred above.
The sound speed of the ionized medium is $C_{\rm II}=\sqrt{\gamma k
  T_e/m_p}$, where $m_p$ is the mass of the proton and $k$ is the
Boltzmann constant. Assuming the adiabatic index $\gamma=\frac{5}{3}$
and an average $T_e=4900$\,K, we found $C_{\rm II}\approx
8.2$\,km~s$^{-1}$. Therefore, the dynamic age for the ultracompact
H$_{\rm II}$ region W51e2-W is about $5000$\,yr, which appears to be a
few times older than the outflow from W51e2-E (see $t_{\rm out}$ in
Table \ref{tab:CO}).

High-resolution observations of absorption spectra of the HCN(4-3)
from W51e2-E and W51e2-W show that at present W51e2-E dominates the
gas accretion from the molecular core W51e2 (see the HCN(4-3)
absorption region, blue color in Figure \ref{fig:RGB}). If the
accretion rate is a constant of 1.3$\times10^{-3}$ M$_{\sun}$
yr$^{-3}$ \citep{SZH10}, W51e2-E would have accreted a total of
$\sim$6 M$_{\sun}$ over the 5000\,yr period, $<$5\% of the total mass
(140 M$_{\sun}$; Shi et al. 2010) of the sub-core. Thus, most of the
mass accumulation for W51e2-E probably occurred before the star(s)
formed in the ultracompact H$_{\rm II}$ region W51e2-W. After an O8
star or a cluster of B type stars formed in W51e2-W, the overwhelming
radiation pressure and strong turbulence in W51e2-W halted the
accretion in the sub-core region. Therefore, W51e2-E might speed up
its formation of stars since W51e2-W is no longer competing for
accretion.

In summary, a bipolar molecular outflow in W51e2 is confirmed and the
massive protostellar core W51e2-E is identified to be the origin of
this powerful outflow. In comparison to the ultracompact H$_{\rm II}$
region W51e2-W, W51e2-E appears to be the dominant accretion source at
present, where the active formation of massive stars takes place.

\acknowledgments
We thank Prof. Ed Churchwell and the referee for helpful comments.
H.S. and J.L.H. are supported by the National Natural Science
Foundation (NNSF) of China (10773016, 10821061, and 10833003) and the
National Key Basic Research Science Foundation of China
(2007CB815403).

\end{document}